\documentclass[%
 reprint,
superscriptaddress,
 amsmath,amssymb,
 aps,
]{revtex4-1}

\usepackage{graphicx}
\usepackage{dcolumn}
\usepackage{mathrsfs}
\usepackage{bm}
\usepackage{xcolor}

\newcommand{\cmmnt}[1]{}

\newcommand{\beginsupplement}{%
        \setcounter{table}{0}
        \renewcommand{\thetable}{S\arabic{table}}%
        \setcounter{figure}{0}
        \renewcommand{\thefigure}{S\arabic{figure}}%
     }

\begin{document}

\preprint{APS/123-QED}

\title{Wrapping of Microparticles by Floppy Lipid  Vesicles}

\author{Hendrik T. Spanke}
\affiliation{ETH Z\"{u}rich}%
\author{ Robert W. Style}%
\affiliation{ETH Z\"{u}rich}%
\author{Claire Fran\c{c}ois-Martin}%
\affiliation{ETH Z\"{u}rich}%
\author{Maria Feofilova}%
\affiliation{ETH Z\"{u}rich}%
\author{Manuel Eisentraut}%
\affiliation{Department of Physics, University of Bayreuth}%
\author{Holger Kress}%
\affiliation{Department of Physics, University of Bayreuth}%
\author{Jaime Agudo-Canalejo}
\affiliation{Max Planck Institute for Dynamics and Self-Organization (MPIDS)}
\author{Eric R. Dufresne}
\email{eric.dufresne@mat.ethz.ch}%
\affiliation{ETH Z\"{u}rich}%

\date{\today}

\begin{abstract}
Lipid membranes, the barrier defining living cells and many of their sub-compartments, bind to a wide variety of nano- and micro-meter sized objects.
In the presence of strong adhesive forces, membranes can strongly deform  and wrap the particles, an essential step in crossing the membrane for a variety of healthy and disease-related processes. 
A large body of theoretical and numerical work has focused on identifying the physical properties that underly wrapping.
Using a model system of micron-sized colloidal particles and giant unilamellar lipid vesicles with tunable adhesive forces, we measure a wrapping phase diagram and make quantitative comparisons to theoretical models.
Our data is consistent with a model of membrane-particle interactions accounting for
 the adhesive energy per unit area, membrane bending rigidity, particle size,  and vesicle radius.

\end{abstract}

\maketitle

The interaction of nano- and micro-objects  with lipid membranes plays an important role in many biological processes.
Examples range from the disease-related entry of viruses and bacteria into cells \cite{Flannagan2012,Dasgupta2014} to healthy docking and priming during vesicular trafficking \cite{Sudhof2013}.
The adhesion of membranes to curvature-stabilizing proteins, such as the BAR family, plays a central role in many membrane-shaping processes of eukaryotic cells \cite{Itoh2005,Kamioka2004,McMahon2005,Gallop2005,Frost2008}.
Finally, nano- and micro-particles can bind to membranes, acting as potential vectors for drug delivery \cite{Dasgupta2017}.

The interaction of particles with membranes therefore has far-reaching consequences in biology and medicine.
This has motivated a rich body of theoretical and computational physical models of membrane-particle interactions \cite{Lipowsky1998,Bahrami2014,Dasgupta2014a,Reynwar2007,Bahrami2012,Raatz2014, Xiong2017,Koltover1999,ruiz2012,Saric2012, Saric2013, Vahid2017}.
Some of the most basic questions revolve around the adhesion of  individual particles with  bilayer membranes.
The simplest theory addressing this question considers the interaction of a spherical particle, of radius $R_P$, with an initially flat membrane connected to a constant tension reservoir \cite{Deserno2004}.
Attractive forces driving adhesion are assumed to be short-ranged, and are  quantified by the adhesive energy per unit area, $\omega$.
Positive adhesion energies drive the membrane to wrap the particle. 
On the other hand, membrane deformation is resisted by its  bending rigidity, $\kappa_b$, and  tension, $\sigma$.
These two membrane properties can be combined to create an important \emph{bendocapillary} lengthscale, $\lambda_\sigma=\sqrt{\kappa_b/\sigma}$ \cite{Deserno2004}.
At lengthscales smaller than $\lambda_\sigma$, membrane deformations are primarily resisted by bending energy, while at longer lengthscales, deformations are mainly opposed by tension.
For low tension or small particles ($R_P \ll \lambda_\sigma$),  wrapping is therefore governed by a balance of only adhesion and bending energy, captured by a second lengthscale, $\lambda_\omega=\sqrt{2\kappa_b/\omega}$.
In that case, membranes should spontaneously wrap particles whenever $R_P>\lambda_\omega$ \cite{Lipowsky1998,Deserno2004}.
As tension increases,  wrapping may require external forces for activation, and wrapped particles may partially unwrap or totally unbind from the membrane.

The assumption of a constant tension reservoir can break down when vesicles are sufficiently deflated, and have enough excess area to wrap a particle without fluctuations being hindered or the membrane being stretched.
In that case, tension no longer plays a role and a pure competition between only bending and adhesion is recovered. 
Recent theoretical studies have shown that  finite curvature of the membrane can be important for particle wrapping in this limit \cite{Agudo-Canalejo2015, Bahrami2016}.

Recent experiments have begun to explore membrane-particle interactions, reviewed in \cite{Idema2019}.
While elucidating a range of higher-order phenomena, these experiments have not tested basic theories of adhesion.
 Experiments have either employed extremely strong irreversible interactions between particles and membranes \cite{Dietrich1997,Wel2016,Sarfati2016,Wel2017}, operated with very tense membranes where there is no significant membrane deformation at the single-particle scale \cite{Dinsmore1998,Wang2019}, or worked with nanoparticles where the interaction between individual particles and membranes cannot be resolved \cite{zuraw2019}.

In this Letter, we experimentally investigate  the wrapping  of micron-sized particles by giant unilamellar vesicles (GUVs) in the biologically-relevant limit of low membrane tension and weak reversible adhesion.
The interaction of particles and membranes is tuned continuously using the depletion effect.
We observe three regimes of interaction between particles and membranes:  non-wrapping, spontaneous wrapping, and activated wrapping.
In the latter case, an external force is required to drive a particle from an unwrapped state to its equilibrium wrapped state.
Detailed comparison with theory suggests an essential role for membrane curvature.

\begin{figure}[t]
\includegraphics[width=0.9\columnwidth]{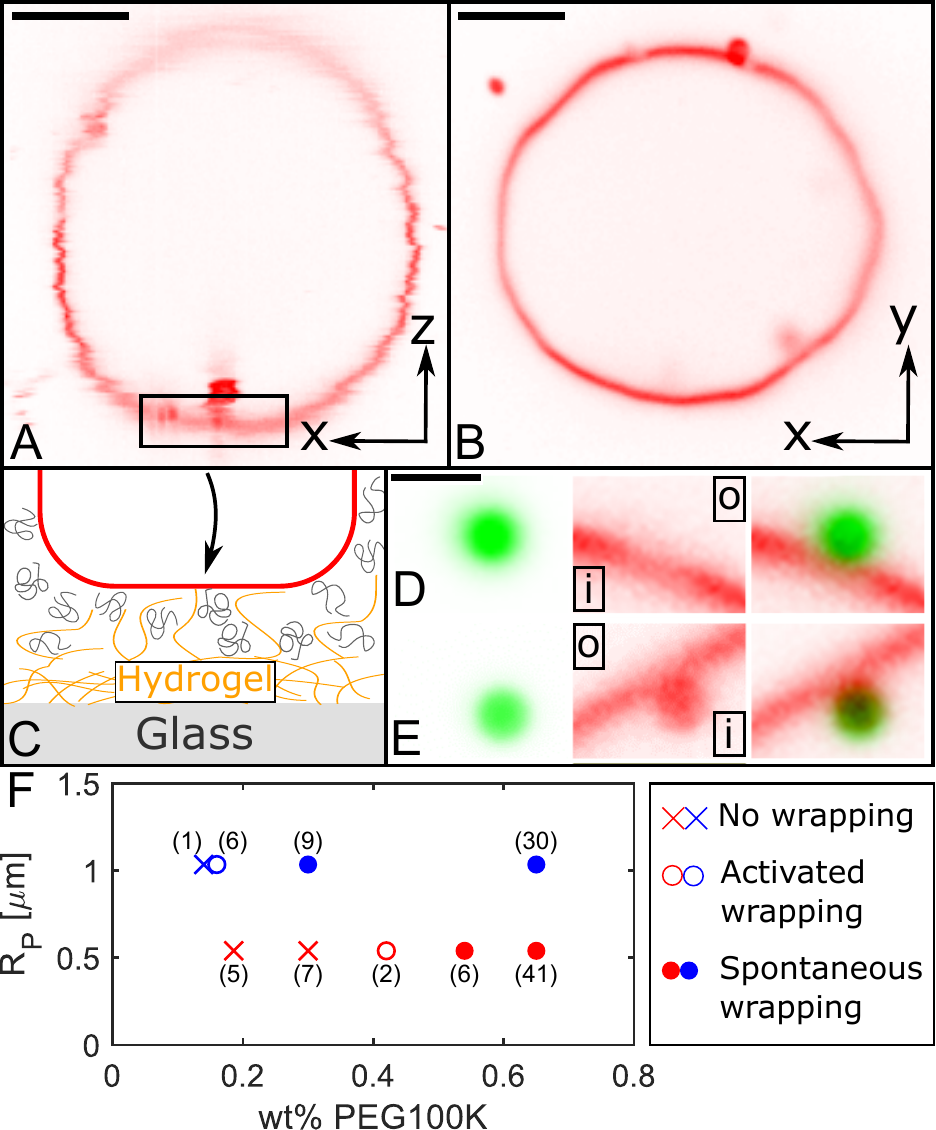}
\caption{\label{fig:system} \emph{Model system}.  \emph{(A,B)} cross-sections of a confocal image stack of a GUV sedimented on a PEG-DA hydrogel featuring obvious thermal fluctuations, despite the presence of $0.65$ wt\% PEG. Scale bars are $10\ \mu $m in length. \emph{(C)} Schematic demonstrating a hypothesized mechanism for reduced depletion interactions against a hydrogel. \emph{(D,E)}  Confocal images of $1.1~\mathrm{\mu m}$ PS particles (green) and POPC membranes (red). The inside and outside of the GUV are indicated by i and o respectively.  The particle in (D) with 0.24 wt\% PEG100K does not deform the membrane. The membrane wraps the particle in (E) at  0.53 wt\% PEG100K. The scale bar is $2\ \mu m$ in length. \emph{(F)} Empirical phase diagram based on the particle radius and amount of PEG depletant in the system. Numbers next to each data point indicate the number of membrane-particle pairs that were probed in each condition.
 }
\end{figure}

Our model system consists of a dispersion of micron-sized polystyrene particles ($1.08 \pm 0.04\ \mathrm{\mu m}$ and $2.07 \pm 0.03\ \mathrm{\mu m}$ in diameter) and  GUVs in polymer solutions.
The GUVs, consisting of 1-palmitoyl-2-oleyl-sn-glycero-3-phosphocholine (POPC) with 1\% 1,2-dioleoyl-sn-glycero-3-phosphoethanolamine-N-(lissamine rhodamine B sulfonyl) (Rhodamine PE), are made by electroformation in a 280 mOsm/kg sucrose solution \cite{Angelova1986,Angelova1992,dimova2019}.
The osmolality of the solvent is adjusted  through the addition of glucose (approximately 270 mM) to a slightly hypertonic value of 290 mOsm/kg.
Over the course of hours, this slight osmotic imbalance drives the deflation of vesicles, leading to very low tensions, demonstrated in later sections.

To achieve tunable weak adhesion between particles and GUVs, we employ depletion interactions \cite{Dinsmore1998}.
Generally, the depletion interaction between two objects has the form, $E_{ad} = \Pi \Delta V$ \cite{Asakura1958,Vrij1976}, where $\Pi$ is the osmotic pressure of the depletant and $\Delta V$ is the change of the depletant's excluded volume due to contact.
For low concentrations, $n$, of depletant, the osmotic pressure is well approximated with the ideal form,
$\Pi=nk_BT$.
The excluded volume,  $\Delta V =-A_{co} \ell$,  where, $\ell$ is the range of range of the depletion interaction, and $A_{co}$ is the contact area (\emph{i.e.} the area over which the two surfaces are within depletion range of each other).
Thus, the adhesion energy density,  $\omega=-E_{ad}/A_{co}$,  has the form
\begin{equation} \label{eq:depletion}
    \omega=n \ell k_{B}T 
\end{equation}
For hard sphere depletants, $\ell$ is expected to be equal to their diameter.
For polymer depletants in a good solvent, $\ell \approx R_g$ \cite{Tuinier2001}.
As the depletion agent, we use polyethylene glycol (PEG) with a molecular weight of $10^5$ g/mol, which has a radius of gyration $R_g$ of about 16 nm, and an overlap concentration of 0.99 wt\% \cite{Ziebacz2011, rubinstein2003polymer}.
We used a range of PEG concentrations between 0.14-0.65 wt\% ($\pm$ 0.016 wt\%) in the samples, yielding adhesion energies from 0.6 to 2.6 $\mathrm{\mu J/m^2}$.
In this range, micron-sized particles can strongly bind membranes while still enjoying reversible interactions with each of its consitutive lipid molecules.

The main challenge in using depletion interactions for studies of particle-vesicle adhesion is their non-specificity. 
Depletion forces not only drive adhesion of particles to vesicles, but also the adhesion of vesicles to the surface of the sample chamber.
At the depletion strengths used here, vesicles spread on flat glass surfaces. 
This significantly increases their tension, and usually leads to rupture  \cite{Richter2006}, as shown in Supplemental Fig.~S1.
To suppress adhesion between vesicles and the walls of the sample chamber, we coat it with a loose network of poly(ethylene glycol) diacrylate (PEG-DA), described in the Supplement.
On this surface, adhesion is strongly reduced and  vesicles remain floppy, even after sedimenting against the surface.
An example of such a GUV, imaged with a confocal microscope is shown in Fig.~\ref{fig:system}AB.
As shown in Fig.~\ref{fig:system}C, we hypothesize that the PEG-DA network is permeable to the depletion agent, reducing its effect.

With a robust mechanism for controlling the adhesion energy, we can now determine which conditions lead to the wrapping of particles by the membrane.
The state of wrapping is easily inferred from fluorescent images of the particle and membrane.  
Fig.~\ref{fig:system}D shows confocal micrographs of a $1.08~\mathrm{\mu m}$  diameter fluorescent polystyrene particle in proximity to a fluorescently tagged GUV. 
In this case, the particle is `unwrapped': the center of mass of the particle remains outside the convex hull of the GUV and there is no significant membrane deformation.  
By contrast, Fig.~\ref{fig:system}E shows a `wrapped'  particle. 
Not only has the particle been pulled to the other side of the membrane's convex hull,  but the membrane is strongly deformed and covers a large portion of the particle surface.

An empirical phase diagram showing the  dependence of wrapping behavior on particle size and polymer concentration is shown in Fig.~\ref{fig:system}F.
To efficiently explore the interactions while minimizing unobserved membrane-particle binding events, we worked at low particle volume fractions $\phi<10^{-5}$ and used optical tweezers to bring particles close to the GUV surface.
At high polymer concentrations, we observe \emph{spontaneous wrapping} of particles by membranes, shown by the filled circles in Fig.~\ref{fig:system}F  (see supplemental movie S1).
At intermediate polymer concentrations, we observe \emph{activated wrapping}, indicated by the open circles in Fig.~\ref{fig:system}F. 
In this case, releasing a particle close to a GUV was insufficient to induce wrapping. 
Instead, wrapping could only be initiated by pushing the particle against the GUV with the optical tweezers.
Nevertheless, particles remain stably wrapped after the laser was turned off (see Supplemental movie S2).
At low polymer concentrations, there is \emph{no wrapping}, as indicated by the x's in Fig.~\ref{fig:system}F.
In these cases, we can force the membrane to wrap a particle using optical tweezers.
However, as soon as the trap is turned off, the membrane returns to its initial state and the particle diffuses away (see supplemental movie S3). 
Note that the minimal polymer concentration needed to drive wrapping increases as the particle size decreases.
This is qualitatively consistent with a simple competition of adhesion and bending rigidity, as summarized in the introduction.
In that picture, activated wrapping should only occur at finite membrane tension \cite{Deserno2004}.

\begin{figure}[]
\includegraphics[width=0.9\columnwidth]{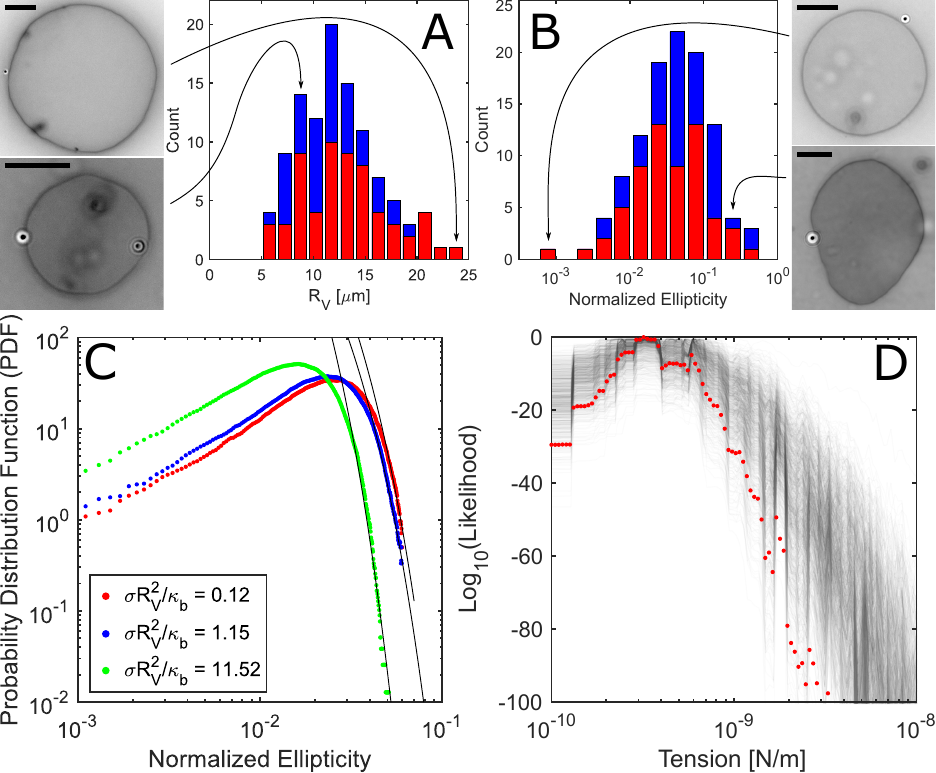}
\caption{\label{fig:membrane} \emph{Geometry and tension of GUVs}
\emph{(A,B)} Histograms of vesicle radii $R_V$, and normalized ellipticities.
GUVs wrapping $0.54$ and $1.04 \ \mu $m radii particles are indicated in red and blue, respectively.
The micrographs show examples of GUVs just before a spontaneous wrapping event.
The scale bars are $10\ \mu m$ in length.
\emph{(C)} Histogram of simulated ellipticities for reduced tensions, $\sigma R_V^2 / \kappa_b=0.12, 1.15, 11.52$. 
Black lines are exponential fits to the high-ellipticity tails. 
\emph{(D)} Logarithm of the likelihood for a range of membrane tensions for the entire data set (red) and for 1000 randomly selected sub-sets of the data (light gray), each using half of the data set.
Values have been shifted so that the most likely tension of each data set has a value of zero.
}
\end{figure} 

The tension and bending rigidity of lipid bilayer membranes of individual vesicles, with radii $R_V$,  can be extracted by an analysis of their thermal shape fluctuations of GUVs \cite{Loftus2013,Engelhardt1985,Faucon1989}.
For moderately tense vesicles, where $\lambda_\sigma \approx 0.1 R_V$, both the tension and bending rigidity can be reliably determined by comparing the mean amplitudes of the Fourier fluctuation modes to the expected Boltzmann distribution.
$\lambda_\sigma$ needs to be sufficiently small to ensure a reasonable number of fluctuation modes with a wavelength $\lambda>\lambda_\sigma$.
This approach is called  vesicle fluctuation analysis (VFA) and is summarized in the Supplement. 
Application of VFA to individual vesicles under the conditions in our experiments consistently report $\kappa_b= 33 \pm 8\ k_BT$ as shown in Supplemental Figure S3.
The resulting tensions are consistent with zero tension, with uncertainties varying from $10^{-9}$ to $10^{-7}~\mathrm{N/m}$, shown in Supplemental Figure S3.

To efficiently place an upper limit on the membrane tension for the vesicles used in the adhesion experiments,   we inferred the most likely tension of the ensemble from simple measures of vesicle shape.
We extracted the major and minor axes ($a$ and $b$) of each GUV, just before it came into contact with the particle.
Histograms of the mean vesicle radii, $R_V=(a/2+b/2)/2$, and the normalized ellipticity $(a-b)/(a+b)$, are shown in Fig.~\ref{fig:membrane}AB.
The vesicles' mean radii range from 5 to 25 $\mathrm{\mu m}$ and the normalized ellipicity varies from $10^{-3}$ to $0.55$.
Using a Monte Carlo simulation of vesicle fluctuations near equilibrium based on the same assumptions of VFA, described in the Supplement, we calculated the probability distribution of the normalized ellipticities for a range of membrane tensions. 
Three probability distributions are shown in Fig.~\ref{fig:membrane}C.
The likelihood of each tension is determined by multiplying the probabilities of all the experimentally observed ellipticities.
The log-likelihood of the tensions between $10^{-10}$ and $10^{-8}$ N/m are shown in Fig.~\ref{fig:membrane}D.
A tension of $3.2\times10^{-10}$ N/m is the most likely tension to describe all GUVs observed. 

The VFA and maximum likelihood results show that tensions are very low, compatible with zero and comparable to the bending scale $\kappa/R_V^2 \approx 1.25\times10^{-9}$ N/m, suggesting that our vesicles are in the deflated, floppy regime. This is apparent from the strong deviations from sphericity observed (Fig.~\ref{fig:membrane}B, Movie S2). As described in the introduction, in this regime the concept of an effective tension ceases to be useful and the system can be regarded as tensionless.


\begin{figure}[]
\includegraphics[width=0.8\columnwidth]{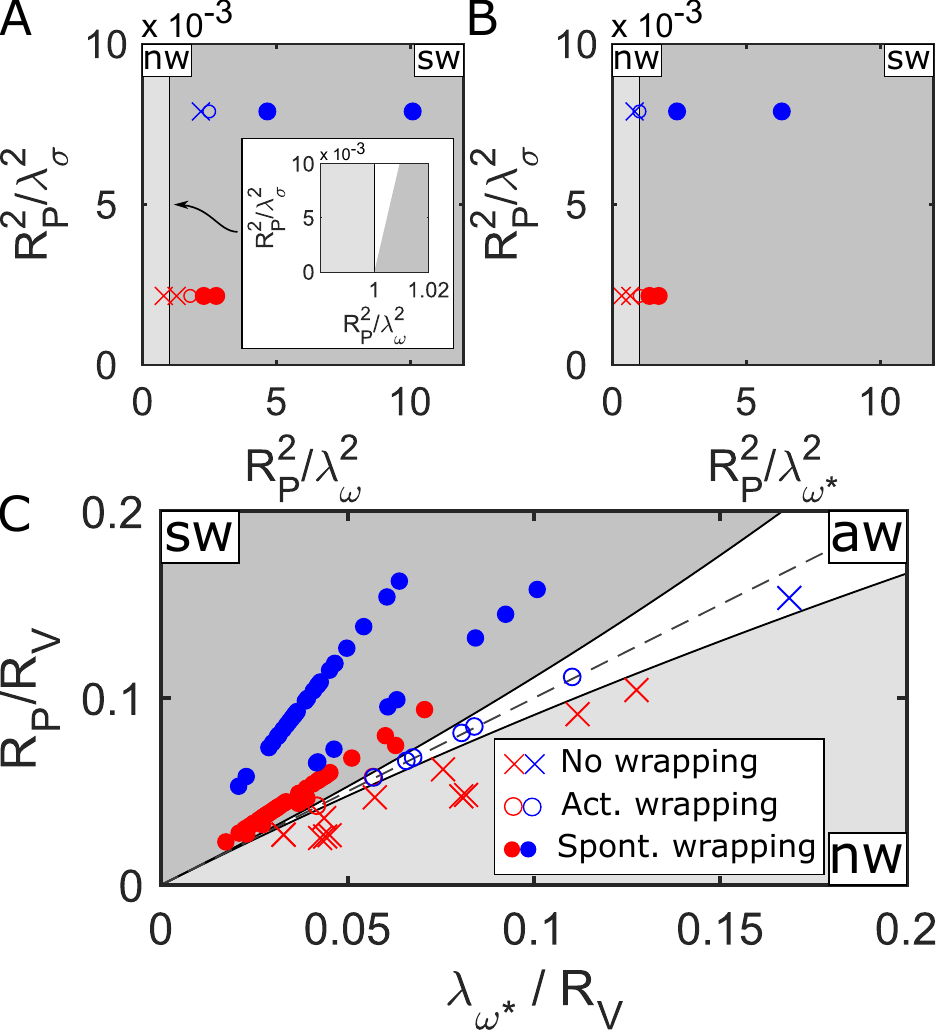}
\caption{\emph{Comparison of experiment and theory}  \emph{(A)} Phase diagram for flat-membranes \cite{Deserno2004} compared to data from Fig.~\ref{fig:system}F using a membrane tension \cmmnt{corresponding to the thermodynamic lower limit of $\kappa/R_V^2$} of $10^{-9}$ N/m and adhesion energy density $\omega$ from depletion (Eq. \ref{eq:depletion}). Inset shows the narrow region of activated wrapping.  Shading indicates theoretical predictions:  dark gray for spontaneous wrapping (sw), light gray for non-wrapping (nw), and white for activated wrapping (aw).
\emph{(B)} same with corrected adhesion energy density $\omega^{*}$ incorporating steric interactions (Eq. \ref{eq:depletion2} and $c=0.04$).
\emph{(C)} Phase diagram for curved membranes  \cite{Agudo-Canalejo2015} using the corrected adhesion energy density $\omega^{*}$ of Eq. \ref{eq:depletion2} with $c=0.04$.
}
\label{fig:theory}
\end{figure}
 
Indeed, using $10^{-9}$ N/m as a representative value of tension, plotting our results in the theoretical phase diagram of \cite{Deserno2004} shows that, at such low tensions, the predictions are indistinguishable from those for a tensionless membrane (Fig.~\ref{fig:theory}A). The region of activated wrapping exists only in a tiny band around the boundary $R_P^2/\lambda_\omega^2=1$, see the inset of Fig.~\ref{fig:theory}A, corresponding to a difference in PEG concentration of 0.0023 wt\%, much smaller than our precision in defining the polymer concentration.

Furthermore, we find that we consistently need higher adhesion energies for spontaneous wrapping than predicted by the theory. This shift can be understood as a consequence of thermal fluctuations.
Shape fluctuations increase the range of steric repulsions with a membrane,  scaling like $1/x^2$, where $x$ is the surface to surface separation \cite{Helfrich1984}.
For $x<\ell$, the net energy per unit area is
\begin{equation}
    E(x)/A =-n k_BT (\ell-x) + c\frac{(k_{B}T)^2}{\kappa_b x^2}.
    \label{eq:depletion2}
\end{equation}
The first term captures the  separation dependence of the depletion force.
The second term is the steric repulsion and features an unspecified dimensionless constant $c$, predicted to be in the range of 0.01-0.23 \cite{Helfrich1984, Dinsmore1998}.
Minimizing with respect to the separation, we find a reduced adhesion energy,
\begin{equation} \label{adhesioneq}
    \omega^{*}=nk_{B}T \left[ \ell - 3 \left( \frac{ck_{B}T}{4n\kappa_b}\right)^{1/3} \right]
\end{equation}
Accounting for this additional repulsive interaction, we find a good fit between theory and experiments for  the  transition from free particles to activated wrapping  using a value of $c=0.04$ for both particle sizes, as shown in Fig.~\ref{fig:theory}B.
Theory and experiment are consistent for a range of $c$ from 0.028 to 0.055 (see Supplement).
These $c$ values are also consistent with 
consistent with previous Monte Carlo simulations of membrane-solid wall repulsion,  \cite{Netz1995,Gompper1989,Janke1989}.


However, this correction for thermal fluctuations does not address the presence of a robust activated-wrapping regime.
To understand this, we turn to recent theoretical advances on particle wrapping by deflated vesicles, for which membrane area and volume are conserved and tension does not play a role \cite{Agudo-Canalejo2015, Bahrami2016}.
There, activated wrapping can occur when the membrane curves away from the particle at their point of initial contact, such as when a particle attaches to a GUV from the outside.
The corresponding phase diagram is shown in Fig.~\ref{fig:theory}C.
It is spanned by two variables, the relative size of the particle and the vesicle ($R_P/R_V$), and the relative size of the adhesion lengthscale and the vesicle ($\lambda_\omega/R_V$).
The transition from no wrapping (light gray) to activated wrapping (white) occurs at $R_P/R_V=1/(1+(\lambda_\omega/R_V)^{-1})$.
The transition from activated wrapping to spontaneous wrapping (dark gray) occurs at larger particles sizes, $R_P/R_V=1/((\lambda_\omega/R_V)^{-1}-1)$.
As vesicles become more strongly curved,  the two transitions move further apart, broadening the range of adhesion energies where activated wrapping is expected.  
In the limit of low curvatures, the two transitions merge, and reduce to the result for a tensionless planar membrane of \cite{Deserno2004},  $R_P/R_V=\lambda_{\omega}/R_V$, shown here as a dashed line.
Superimposing the data from Fig.~\ref{fig:system}F on top of this phase diagram, we find good agreement.
This theory can be expanded to incorporate non-zero spontaneous curvature,  \cite{Agudo-Canalejo2015}. but this correction is not necessary to describe our data (see Supplement).\\

We have introduced a model system to probe the wrapping of spherical particles by lipid bilayer membranes featuring tunable adhesive interactions and low membrane tensions.
Our experiments agree with theory accounting for the vesicle curvature and weakened depletion interactions due to thermal shape fluctuations.
Our micron-scale experiments not only have clear connections to the interactions of microplastics with living cells \cite{Browne2008,vonMoos2012}, but they are also relevant to nano-scale interactions of proteins and lipid membranes.  
Like the latter case, our experiments are dominated by bending, \emph{i.e.} $R_P \ll \lambda_\sigma$.  
Our experiments also start to probe regimes where the particle size is comparable to but smaller than the membrane radius of curvature, the typical regime for curvature stabilizing proteins.
However, an isotropic sphere is  a poor approximation for most folded proteins. 
Additionally, many proteins do not simply adsorb to the membrane but also anchor themselves with hydrophobic tails.
Despite these limitations,  work on such model systems helps to establish the physical foundations for an understanding of membrane-particle interactions over a wide range of scales.
Future studies should aim to clarify the nature of membrane-mediated particle interactions and the coupling of  adsorption and self-assembly of particles to generate large-scale shape transformations of membranes.\\

We acknowledge helpful conversations with Raphael Sarfati, Patricia Bassereau, Karine Guevorkian, Markus Deserno, Rumiana Dimova and Reinhard Lipowsky as well as funding from grant number 172824 of the Swiss National Science Foundation and the German Research Foundation (DFG) - project number 391977956 - SFB 1357.

\beginsupplement

\appendix

\section{Materials \& Methods}

\subsection{Materials}
1-Palmitoyl-2-oleoyl-sn-glycero-3-phosphocholine (POPC) and 1,2-dioleoyl-sn-glycero-3-phosphoethanolamine-N-(lissamine rhodamine B sulfonyl) (ammonium salt) (Rh-DOPE) were purchased from Avanti Polar Lipids, Inc. (Alabaster, Alabama).
D-(+)-glucose (BioXtra, $\geq$ 99.5\%) and sucrose (BioXtra, $\geq$99.5 \%) were purchased from Sigma Life Science. 
NaCl (ACS reagent, $\geq$99.0\%), poly(ethylene glycol) diacrylate with an average molecular weight $M_n=700$ and chloroform was bought from Sigma-Aldrich.
Ethanol, absolute was purchased from Fisher Chemicals.
3-(Trimethoxysilyl)propyl methacrylate was purchased from TCI (Tokyo Chemical Industry). 
Poly(ethylene oxide) (also called polyethylene glycol, PEG) powder with average Mv of 100,000, 2-Hydroxy-4'-(2-hydroxyethoxy)-2-methylpropiophenone (Irgacure 2959) as well as poly(ethylene glycol)-block-poly(propylene glycol)-block-poly(ethylene glycol) (PEG-PPG-PEG, Pluronic F108, average Mn $\sim$ 14,600) were bought from Aldrich Chemistry.
Fluorescent polystyrene-particles with a diameter of 1.08 $\mathrm{\mu m}$ and 2.07 $\mathrm{\mu m}$  were purchased from Microparticles GmbH (Berlin, Germany).\\

All chemicals were used as received.\\

\subsection{Electroformation of GUVs}
 POPC was used to make giant unilamellar vesicles by electroformation \cite{Angelova1986,Angelova1992}.
 Rhodamine tagged lipids (Rh-DOPE) are added in the low concentration of 1\%.
 50 $\mu L$ of a 1 mM solution of these lipids was deposited on an ITO-plate using a glass syringe (Hamilton).
 A PDMS-Spacer was placed on the ITO-plate and a second ITO -plate is put on top.
 The chamber was filled with a solution of 280 mOsm/kg sucrose and sealed. both ITO-plates are connected electrically to a signal generator (Keysight 33210A).
 The electroformation protocol consists of gentle increase in AC-voltage over 36 minutes hour from 0 to 1.74 V with a fixed frequency of 10 Hz.
 After the voltage reaches 1.74 V it is left for one hour.
 The voltage is then once more increased to 2.03 V and the frequency lowered to 4 Hz for another hour.
 The electroformation chamber is then kept at a temperature of 4 $^{\circ}$C overnight as is.
 Vesicles with varying sizes between a few $\mathrm{\mu m}$ and up to 50 $\mathrm{\mu m}$ are taken out and stored in the 280 mM sucrose solution from the chamber at 4 $^{\circ}$C where they are stable for up to a few weeks.

\subsection{Surface Treatments}
GUVs suspended in the depletion mix  burst on bare glass, as shown in Fig.~S1.
 To prevent this, the substrate for the observation of the GUVs was covered in a weak PEG-DA hydrogel.
In a first step, a coverslip was coated in 3-(Trimethoxysilyl)propyl methacrylate:
a silane solution of 3 $\mathrm{\mu L}$ silanes in 950 $\mathrm{\mu L}$ of a 95Vol\% ethanol in water solution is prepared, vortexed and left for 3 minutes. The coverslip is UV/Ozone treated for 5 minutes before being coated by 30 $\mathrm{\mu L}$ of the silane solution.
The liquid is spread equally over the entire surface of the coverslip.
After 5 minutes the coverslips are submerged in ethanol and gently rubbed with kimwipes.
The coverslips are stored in a dry environment to avoid humidity unbinding the silanes again.\\

To bind a thin layer of a PEG-DA hydrogel on the coverslips a 1 wt\% solution of Irgacure 2959  was prepared and used to make a 20 Vol\% PEG-DA solution.
20 $\mathrm{\mu L}$ of this gelling solution is dropped on the silanized coverslip and a second coverslip treated with Rain-x rain repellent (ITW Global Brands) is placed on top.
The assembly is placed under a UV lamp (VL-8.LM from LTF Labortechnik GmbH \& Co. KG) and irradiated with 0.3 mW/cm$^2$ of UV-light at 365 nm wavelength for 3 minutes.
The thusly produced hydrogel can be stored submerged in water.

\begin{figure}[]
\includegraphics[width=0.4\textwidth]{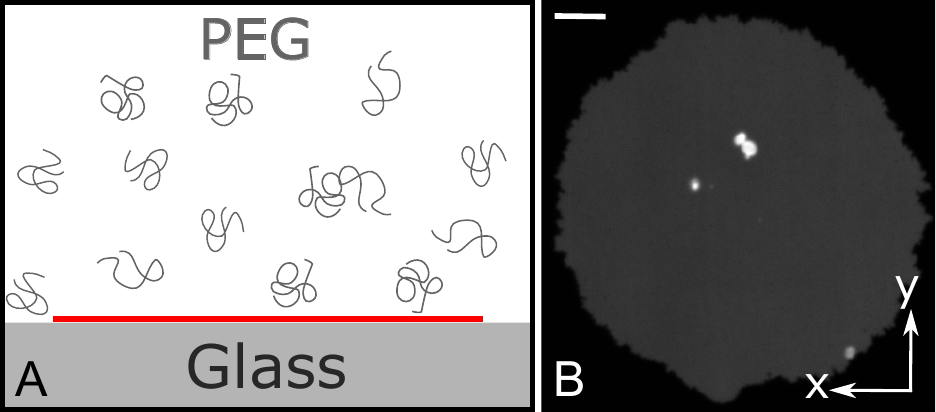}
\caption{\label{FigApp5} \emph{(A)} Sketch of a burst GUV after adhesion to bare glass in the presence of a depletion agent.
\emph{(B)} Confocal image of a patch of lipids on the glass substrate with no hydrogel present.
The scalebar is $10\ \mathrm{\mu m}$ in length.
}
\end{figure}

 \subsection{Preparing the Sample and Assembling the Sample Chamber}
 When preparing a sample chamber the top rain-X treated coverslip is removed from the PEG-DA coated coverslip and the hydrogel briefly rinsed with water.
 An imaging spacer (Grace Bio-Labs SecureSeal™ imaging spacer  purchased through Sigma-Aldrich) was placed on the hydrogel-coated coverslip. To have the spacer stick parts of the hydrogel is cut away to obtain a square patch of similar size to the hole in the spacer.
 The sample volume was then filled with 70 $\mathrm{\mu L}$ depletion medium, 0.3 to 0.5 $\mathrm{\mu L}$ of a 0.025 wt\% particle suspension and 10 $\mathrm{\mu L}$ sucrose solution containing the GUVs.
 The GUVs are suspended in a media introducing depletion interactions in the system.
 
 The depletion medium stock consists of 10 mM NaCl to screen electric charges and limit the Debye length to about 3 nm, 0.05 wt\% of Pluronic F108, a desired amount of polyethylene glycol (PEG) varied between 0.16 and 0.75 wt\% depending on the desired adhesion energy density in the system as well as glucose to have the osmolality add up to 290 mOsm/kg.
 Typically a little less than 270 mM of glucose is needed.
 The amount of Pluronic F108 is kept constant between experiments and is added to passivate the surface of the particles \cite{Phillips2014}.
 
 A second clean cover slip is added on top to seal the sample and prevent evaporation.
 Samples prepared this way showed no significant evaporation over some hours. Samples equilibrating overnight were submerged in the same depletion medium as was present in the sample.

  \subsection{Optical Microscopy and Micromamipulation}
 Confocal imaging was done on a Nikon Ti2 microscope with a 3i Spinning Disk Confocal system using a 60x water immersion objective lens. Images were taken with an Hamamatsu ORCA-Flash 4.0, C13440.\\
 
 Experiments using optical tweezers were done with Nikon Ti Eclipse inverted microscopes using a 60x water immersion objective lens.
 The trapping laser was either a LUXX 785-200 Laser from Omicron Laserage Laserprodukte GmbH with a wavelength of a wavelength of 785 nm and a maximum power of 200 mW (at ETH) or a Ytterbium Fiber Laser from IPG Photonics (model number is YLM-5-LP-SC) with a wavelength of 1064nm and a maximum output power of 5 W (at Bayreuth).
 Videos were taken with a Hamamatsu ORCA-Flash 4.0, C13440.\\

   \subsection{Vesicle Fluctuation Analysis} \label{VFA}
 The implemented analysis of the vesicle fluctuation closely follows \cite{Faucon1989,Meleard2011}.
We limited this analysis to  vesicles that were fluctuating about a spherical shape, with ellipticities ranging from $5.2\times10^{-3}$ to $7.6\times10^{-2}$.
 A video of at least 2400 frames at 50 frames per second is recorded and the contour $\rho(\phi,t)$ is determined in each frame of the acquisition.
 For each pixel on the circumference a radial line scan of the image intensity is extracted.
 The maximum in intensity is fit with a parabolic function to determine the radial contour position with subpixel resolution.
 For each frame the amplitude of the fluctuation is calculated as:
 
 \begin{equation}
     u(\phi)=\frac{\rho(\phi,t)-\langle\rho(t)\rangle}{\langle\rho\rangle}
 \end{equation}
 
 where $\langle\rho(t)\rangle$ is the mean radius of the contour at time t and $\langle\rho\rangle$ is the mean radius over all times and angles.
 This contour amplitude is decomposed into its constituent Fourier modes and we get the mean squared amplitude of each mode through:
 
 \begin{equation}
    \chi_m=\frac{2}{N^2} \left|\tilde{u}_m\right|^2,
 \end{equation}
 where $N$ is the number of measurement points around the vesicle, and $\tilde{u}_m$ is the discrete Fourier transform of $u_n=u(\phi_n)$ (with $\phi_n=2\pi n/N$, $n=0,1,...,N-1$):
 \begin{equation}
      \tilde{u}_m=\sum_{n=0}^{N-1} u_n e^{-i\phi_n m}
 \end{equation}

 The average mean-squared amplitude, $\langle \chi_m\rangle=2/L_m$.
 These values are then fitted with the theoretical expression for a quasi-spherical vesicle of radius R:

   \begin{equation}
    L_m(\kappa/k_bT,\bar{\sigma})=\frac{\kappa}{k_bT}\frac{1}{\sum_{n\geq m}^{n_{max}} \left[Q^m_n(0)\right]^2/\lambda_n(\bar{\sigma})}
 \end{equation}
 to obtain both the membrane tension and the bending rigidity.
 Here, 
    \begin{equation}
    \lambda_n(\bar{\sigma})=(n+2)(n-1)\left[\bar{\sigma}+n(n+1)\right]
 \end{equation}
 and
  \begin{equation} \label{spharm}
    Q^m_n(cos\theta)=(-1)^m \sqrt{\frac{2n+1}{4\pi}\frac{(n-m)!}{(n+m)!}}P^m_n(\cos\theta),
 \end{equation}
 where $\bar{\sigma}=\sigma R^2/\kappa$ is the reduced membrane tension and $P^m_n(\cos\theta)$ the associated Legendre function.
 
\begin{figure}[]
\includegraphics[width=0.475\textwidth]{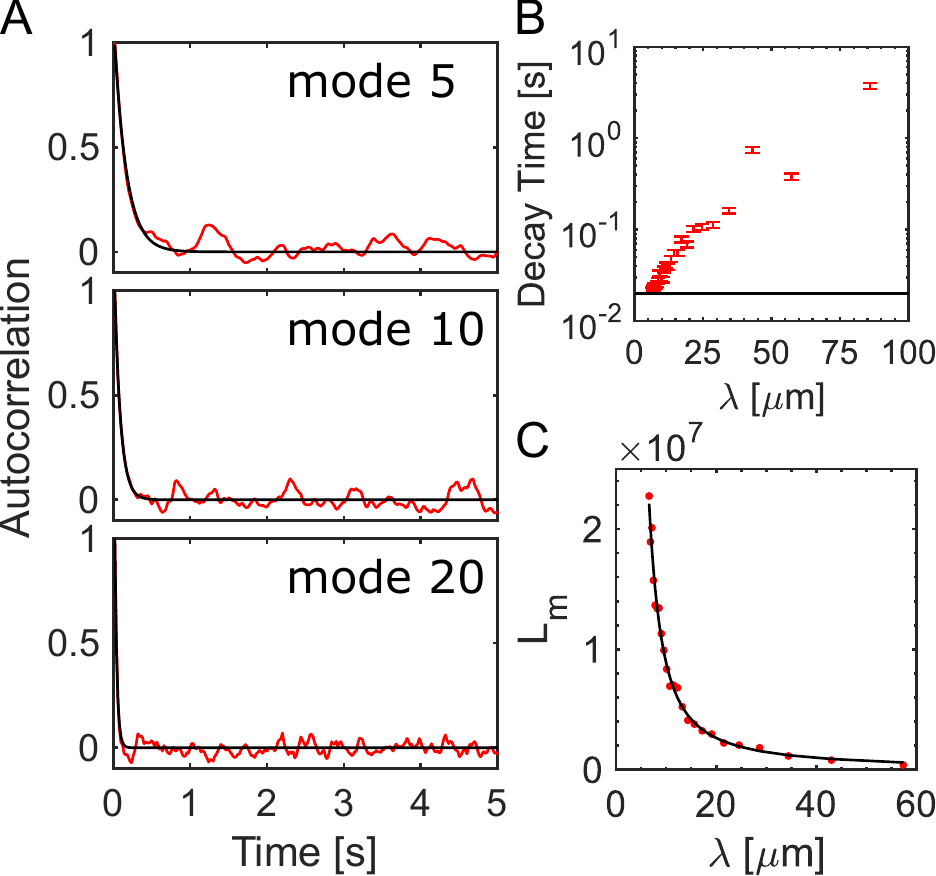}
\caption{\label{FigApp3}  \emph{(A)} Autocorrelations of selected modes 5, 10 and 20
 \emph{(B)} Decay times as a function of mode wavelength $\lambda$. The horzizontal black line indicates the exposure time.
 \emph{(C)} $L_{m}$ values as a function of the mode wavelength (red dots) and the resulting theoretical curve that was fitted.
The result for this GUV is a bending rigidity $\kappa_b$ of $33.6 \pm 4.7$ $\mathrm{k_{B}T}$ and a membrane tension $\sigma$ of $8.5 \pm 2.9 \times 10^{-8}$ N/m.
}
\end{figure}

The decay times of the individual modes allow for an appropriate selection of modes to be analyzed.
The modes must not be decaying on timescales lower the exposure time or they will not be resolved. 
Fig.~S2B shows the decay times and wavelengths of modes $m$ measured in a GUV of 27.37 $\mu$m in radius.
The horizontal line shows the exposure time of the video acquired.
The decay times were obtained by fitting the time autocorrelation of the mean squared amplitude of each mode.
Example time autocorrelations and their fit with an exponential decay function are shown in Fig.~S2A for modes 5, 10 and 20.
The modes shown have decay times and wavelengths of 0.16 s \& 34.39 $\mu$m, 0.078 s \& 17.20\ $\mu$m, 0.027 s \& 8.60 $\mu$m respectively.
The resulting $L_{m}$ values as a function of the wavelength $\lambda$ is shown in Fig.~S2C.\\

The resulting membrane tension, bending rigidity and mean errors for 19 different GUVs are $3.5\times 10^{-8} \pm 2.8\times10^{-8}$ N/m and $32.8\pm 8.4$ $\mathrm{k_{B}T}$ respectively.
The individual measurements and their errors are shown in Fig.~S3AB.\\

 \begin{figure}[]
\includegraphics[width=0.375\textwidth]{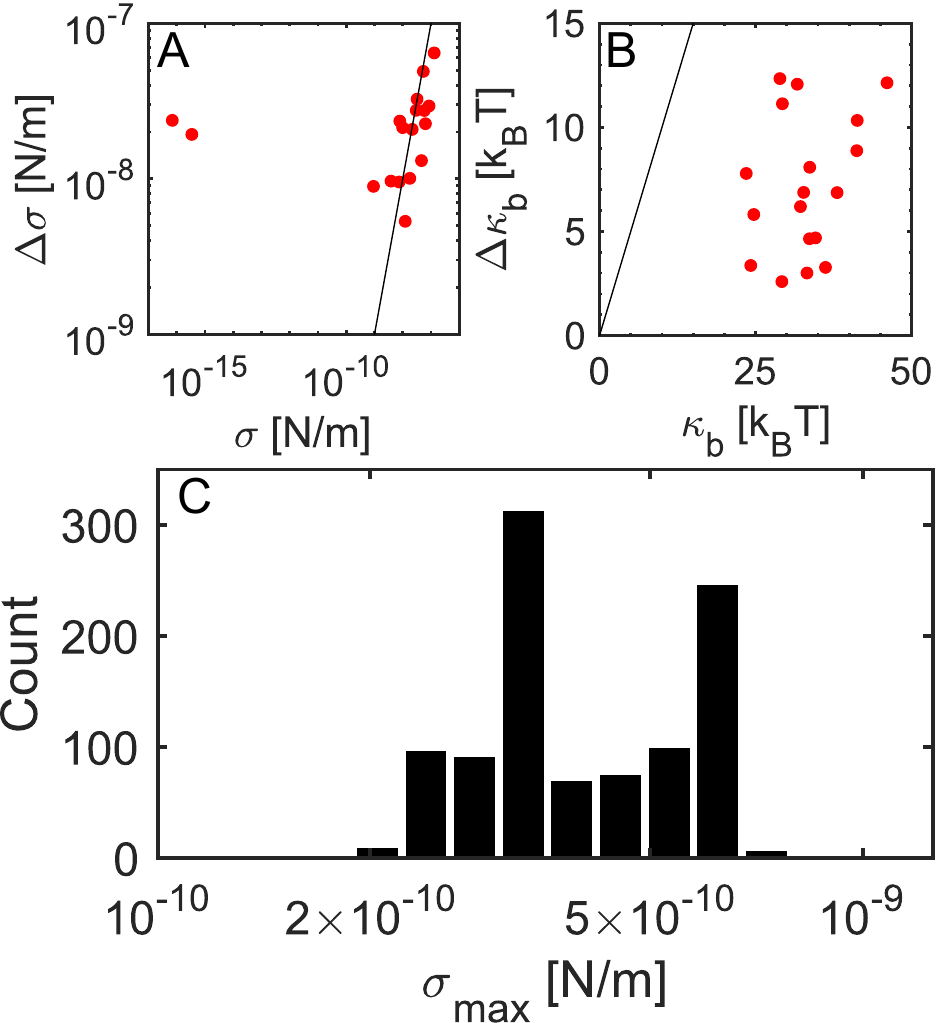}
\caption{\label{FigApp3_2} \emph{(A)} Values for the membrane tension and their error obtained by VFA. The black line indicates an error of 100\%.
\emph{(B)} Values for the membrane bending rigidity and their error obtained by VFA. The black line indicates an error of 100\%.
\emph{(C)} Histogram of the most likely tension in the 1000 randomly selected subsets in Fig.~\ref{fig:membrane}D.
}
\end{figure}

\subsection{Monte Carlo Simulation of Vesicle Fluctuations}
The method used to create vesicle contours with a given radius $R_V$, bending rigidity $\kappa$ and membrane tensions $\sigma$ is based on the same theoretical background as VFA \cite{Faucon1989}.\\

The shape of a vesicle at time $t$ is given by
\begin{equation}
    R(\theta,\phi,t)=R_V(1+\delta u(\theta,\phi,t)),
    \label{eqn:R}
\end{equation}
where the perturbations to a sphere are given by\begin{equation}
    \delta u(\theta,\phi, t)=\sum_{n=0}^{n_{max}} \sum_{m=-n}^{m=+n} U_n^m(t) Y_n^m(\theta,\phi)
\end{equation} 
where $\theta$ and $\phi$ are the spherical polar angle coordinates, $t$ is a timepoint, $n_{max}$ is the maximum number of modes considered, and $Y_n^m(\phi,\theta)$ are Laplace's spherical harmonic functions.

We determine the amplitudes of all modes, $U_n^m(t)$, with $n>1$ by noting that each mode is independent, and on average has an energy of $k_bT/2$ (from the equipartition theorem), and follows Boltzmann statistics \cite{Meleard2011}.
Thus,
\begin{equation}
    U_n^m(t)	=f(t)\left(
    \frac{2E(t)}{\kappa(n-1)(n+2)[\bar{\sigma}+n(n+1)]}\right)^{1/2},
\end{equation}
where $f(t)$ is randomly either plus or minus one with equal likelihood, and $E(t)$ is a Boltzmann distribution, with average energy $k_BT/2$.
$U_0^0(t)$ is subsequently fixed by requiring the total volume of the vesicle to be constant, and 
all values of $U_1^m(t)$ are set to zero as mode 1 corresponds to translations of the whole vesicle without any shape change.
Finally, we obtain the simulated contour in the equatorial plane of the vesicle from equation (\ref{eqn:R}), by setting $\theta$ to $\pi/2$.
The obtained individual contours $R(\phi,t)$ are completely uncorrelated.\\

For each simulated contour we identify the longest axis as the major axis $a$.
The axis perpendicular to to this major axis is then taken as the minor axis $b$.
We create 10'000 contours for 1000 values of $\sigma R_V^2/\kappa_b$ each over a range covering all experimentally observed GUV sizes and expected membrane tensions.
the simulated normalized ellipticities are binned and form the probability distribution functions shown in Fig.~\ref{fig:membrane}C.

To estimate the uncertainty in the maximum likelihood approach used to determine an upper limit to the membrane tension present in the experiments, we formed random subsets of the the GUVs used by randomly selecting half of all experimental datasets a total of 1000 times.
The maximum likelihood approach is run on each of these subsets and yields the likelihood over the same range of tensions.
Fig.~S3C shows a histogram of the most likely tensions obtained this way.
The most likely tension consistently has a value between $2\times10^{-10}$ and $8\times10^{-10}$ N/m.  



\subsection{Influence of Spontaneous Curvature}
Spontaneous curvature, $m$, can play a role in the wrapping process.
Such spontaneous curvatures of membranes can arise whenever there are asymmetries in solute concentrations inside and outside of the membrane \cite{Lipowsky1998}, as is the case in our system.
The effect of spontaneous curvature can be seen by incorporating it into the phase diagram in Fig.~\ref{fig:theory}C, leading to a modified activated-wrapping/no-wrapping phase boundary \cite{Agudo-Canalejo2015}: \begin{equation} R_P/R_V=1/(1+(\lambda_\omega/R_V)^{-1}-2m R_V).
\label{eq:m}
\end{equation}
Importantly, we see from the equation above, that spontaneous curvature only plays a significant role when $m \gtrsim 1/R_V$.
\emph{i.e.} the wrapping process should be unaffected by spontaneous curvature, unless it is large in comparison to the curvature of the vesicle. In our case, unfortunately it is difficult to measure $m$, and as there are multiple asymmetries across the membrane (sucrose, glucose, NaCl, PEG and Pluronic F108), it is difficult to estimate this with a first-principles calculation.
However, we can use two separate approaches to put limits on its magnitude.

First, we use the technique of \cite{Liu2016, Karimi2018}, of counting the number and direction of tube-like features in all the GUVs used for this study to get a rough estimate of the spontaneous curvature.
We found 70.2\% having none, 10.7\% having tubes on the inside, 14.3\% on the outside and 4.8\% on both sides.  This is consistent with low spontaneous curvatures (with magnitudes not larger than the vesicle curvature) for a majority of the vesicles used in the study.

\begin{figure}[]
\includegraphics[width=0.375\textwidth]{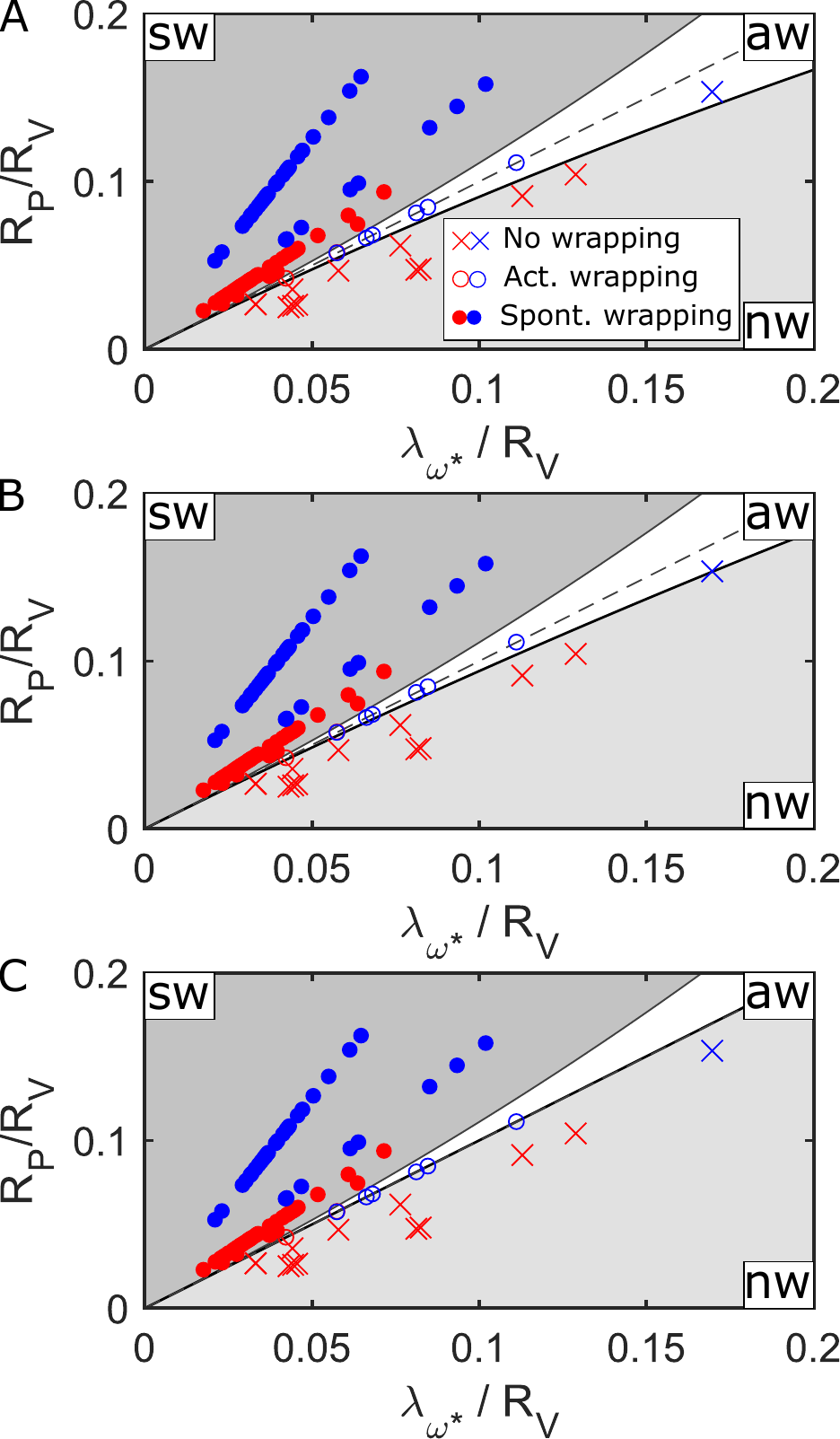}
\caption{\label{FigApp7} Phase diagram from Fig.~\ref{fig:theory}C with the phase boundary between non-wrapping (nw) and activated wrapping (aw) drawn in black for three values of spontaneous curvature $m$.
The spontaneous curvature is set to \emph{(A)} $m=0\ \mathrm{\mu m^{-1}}$, \emph{(B)} $m=0.015\ \mathrm{\mu m^{-1}}$ and \emph{(C)} $m=0.041\ \mathrm{\mu m^{-1}}$
}
\end{figure}

Second, we  estimate the spontaneous curvature by fitting our data to the phase boundary incorporating spontaneous curvature, given by Eq. \ref{eq:m}.
Using $c=0.04$, we vary $m$ and compare our experimental results with the phase boundary between non-wrapping and activated wrapping, as shown in Fig.~\ref{FigApp7}.
When $m=0\ \mathrm{\mu m^{-1}}$, one of non-wrapping points observations is in the activated wrapping regime.
However, for $m$ between 0.015 $\mathrm{\mu m^{-1}}$ (B) and 0.041 $\mathrm{\mu m^{-1}}$ (C), all the experimental data agree with the theoretical predictions.
This suggests that $m\lesssim 0.04 \mathrm{\mu m^{-1}}$.
For reference, $R_V^{-1}=0.09\ \mathrm{\mu m^{-1}}$, so this reinforces the results of the tube analysis that $m<1/R_V$ (meaning that spontaneous curvature effects should be small).

\subsection{Range of Numeric Prefactor to Steric Repulsion of Fluctuations}

\begin{figure}[]
\includegraphics[width=0.375\textwidth]{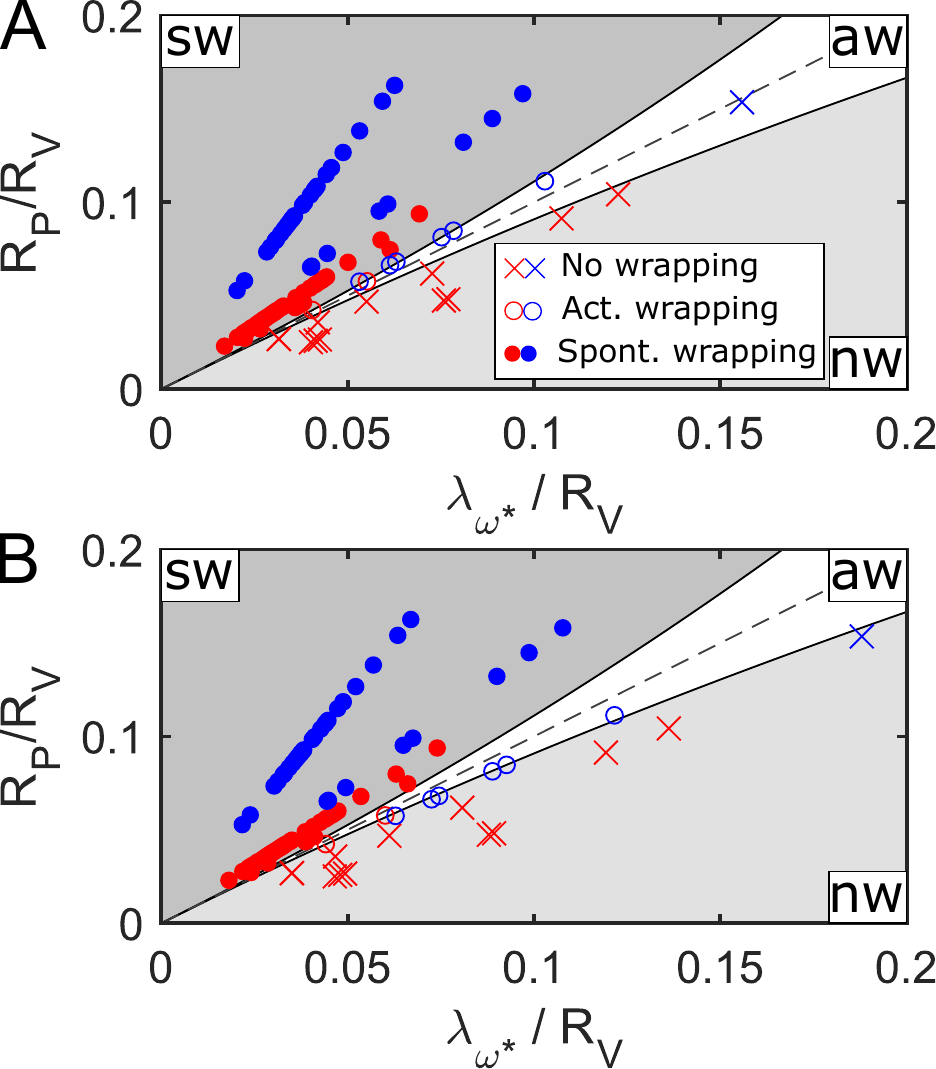}
\caption{\label{FigApp9} Phase diagram from Fig.~\ref{fig:theory}C using the corrected adhesion energy density $\omega^{*}$ of Eq. 2 with (A) $c= 0.028$ and \emph{(B)} $c=0.055$.
}
\end{figure}

The numeric prefactor $c$ to the steric repulsion of fluctuating membranes in Eq. \ref{eq:depletion2}  is found to be in the range of 0.028-0.055 as is shown in Fig.~\ref{FigApp9}.
In this range the proposed phase diagram is able to capture the experimental observations.
A value of 0.04 best aligns the the experimental observations of activated wrapping with the diagonal dashed line describing the very narrow activated wrapping regime for flat membranes.

\subsection{Supplementary Videos}
\begin{itemize}
    \item Movie S1: $2.1~\mathrm{\mu m}$ PS particle undergoing spontaneous wrapping on a GUV with $R_V=13.0\ \mathrm{\mu m}$ and a normalized ellipticity of 0.012.
    The sample contains 0.65 wt\% PEG with $M_w\approx100'000$ g/mol.
    
    \item Movie S2: $2.1~\mathrm{\mu m}$ PS particle undergoing spontaneous wrapping on a GUV with $R_V=18.0\ \mathrm{\mu m}$ and a normalized ellipticity of 0.14.
    The sample contains 0.16 wt\% PEG with $M_w\approx100'000$ g/mol.
    \item Movie S3: $1.1~\mathrm{\mu m}$ PS particle undergoing spontaneous wrapping on a GUV with $R_V=11.2\ \mathrm{\mu m}$ and a normalized ellipticity of 0.06.
    The sample contains 0.19 wt\% PEG with $M_w\approx100'000$ g/mol.
\end{itemize}

\subsection{Supplementary Data}
\emph{ExpDATA.mat:}
MATLAB array with the following columns:
\begin{itemize}
\item Column 1: Particle radius in micrometers
\item Column 2: wt\% of PEG100K in the sample
\item Column 3: 1/2 of the major axis $a$ of the GUV in meters
\item Column 4: 1/2 of the minor axis $b$ of the GUV in meters
\item Column 5: wrapping behavior. 1 - no wrapping, 2 - activated wrapping, 3 - spontaneous wrapping.
\end{itemize} 

\emph{SimDATA.mat:}
MATLAB structure array with the following fields:
\begin{itemize}
\item sigmabar: values of $\sigma R_V/\kappa_b$
\item ellipticity: values of the normalized ellipticity
\item pdf: probability distribution functions for each value of sigmabar over the values of the normalized ellipticity.
\end{itemize}


%

\end{document}